\documentclass{aastex63}
\usepackage[utf8]{inputenc}
\shorttitle{A High-Cadence UV Telescope On The Lunar South Pole}
\shortauthors{Fleming et al.}
\begin{document}

\title{A High-Cadence UV-Optical Telescope Suite On The Lunar South Pole}

\author{Scott W. Fleming}
\affiliation{Space Telescope Science Institute}

\author{Thomas Barclay}
\affil{Goddard Space Flight Center}

\author{Keaton J. Bell}
\affil{University of Washington}

\author{Luciana Bianchi}
\affil{Johns Hopkins University}

\author{C. E. Brasseur}
\affil{Space Telescope Science Institute}

\author{JJ Hermes}
\affil{Boston University}

\author{R. O. Parke Loyd}
\affil{Arizona State University}

\author{Chase Million}
\affil{Million Concepts LLC}

\author{Rachel Osten}
\affil{Space Telescope Science Institute}

\author{Armin Rest}
\affil{Space Telescope Science Institute}

\author{Ryan Ridden-Harper}
\affil{Space Telescope Science Institute}

\author{Joshua Schlieder}
\affil{Goddard Space Flight Center}

\author{Evgenya L. Shkolnik}
\affil{Arizona State University}

\author{Paula Szkody}
\affil{University of Washington}

\author{Brad E. Tucker}
\affil{Mt Stromlo Observatory, the Australian National University}

\author{Michael A. Tucker}
\affil{University of Hawaii}

\author{Allison Youngblood}
\affil{University of Colorado}

\begin{abstract}
We propose a suite of telescopes be deployed as part of the Artemis III human-crewed expedition to the lunar south pole, able to collect wide-field simultaneous far-ultraviolet (UV), near-UV, and optical band images with a fast cadence (10 seconds) of a single part of the sky for several hours continuously. Wide-field, high-cadence monitoring in the optical regime has provided new scientific breakthroughs in the fields of exoplanets, stellar astrophysics, and astronomical transients. Similar observations cannot be made in the UV from within Earth's atmosphere, but are possible from the Moon's surface. The proposed observations will enable studies of atmospheric escape from close-in giant exoplanets, exoplanet magnetospheres, the physics of stellar flare formation, the impact of stellar flares on exoplanet habitability, the internal stellar structure of hot, compact stars, and the early-time evolution of supernovae and novae to better understand their progenitors and formation mechanisms.
\end{abstract}

\section{Introduction}
Space-based telescopes and ground-based surveys using a combination of high-cadence, high-precision photometry with wide fields-of-view have advanced, and in some cases created, many sub-disciplines within astronomy.  \emph{The same principle of wide-field, high-cadence surveys, conducted in the UV, pushes the boundaries of discovery within these same astronomical sub-disciplines in unique ways.}  Such a survey can be achieved using existing technologies: e.g., a modest aperture telescope (30 cm) equipped with CMOS detectors and appropriate protection for lunar dust.  Tracking is not necessary given the slow rotation of the Moon, thus the telescopes can be constructed without moving parts. This project serves as a scientific and technological pathfinder for larger astronomical facilities in a future with increased and sustained human presence on the Moon.

\section{Transiting Exoplanets}
A transiting exoplanet is one that crosses the face of its host star relative to our line of sight.  Their atmospheres (and thus chemical compositions and bulk densities) can be studied via transmission and emission photometry and spectroscopy.  In the UV, the depth of the exoplanet transit can be a tracer of atmospheric mass loss \citep{2003Natur.422..143V} or bowshocks generated by the interaction of the stellar wind and the exoplanet's magnetosphere \citep{2016MNRAS.456.2766A}.  These transit depths can be significantly larger in the UV compared to the optical \citep{2015Natur.522..459E}.  Transits for typical close-in gas and ice giants last a few hours, but the ingress and egress last for only tens of minutes.  High cadence sampling allows for detailed modeling of both the ingress/egress of the transit, and a check for additional variability that may be present from the star's atmosphere itself.  The far ultraviolet (FUV) and near ultraviolet (NUV) ranges contain absorption lines from different species (Hydrogen's Ly $\alpha$, O I, C II in FUV; Mg II in NUV), while simultaneous observations in the optical of the same transit event serves as a crucial benchmark for transit start/stop times and shape comparison.  High-cadence UV observations are also ideal for detecting deep transits of planets orbiting white dwarfs, which last only minutes \citep{2011ApJ...731L..31A}. White dwarfs are the final remnants of the majority of planet hosting stars, and this difficult-to-detect population will reveal the ultimate fate of planets like the Earth.

\section{Stellar Flares}
The formation mechanisms of flares on stars other than the Sun are still poorly constrained from observational data. Because flare timing is random, a long baseline of observation is required to capture a representative sample of flares. The UV is an ideal wavelength regime to study flares because the contrast between a flare and the star's atmosphere is large. Observing a flare in multiple bands simultaneously provides the best constraints on the flare physics, but coordinating multiple telescopes in space and on the ground is difficult for large numbers of targets. Space-based optical telescopes like Kepler and TESS have not been able to resolve short-duration flares. Archival GALEX data allows for sampling at the seconds level, uncovering a population of flares that last 1-5 minutes, but did not continuously observe any single target longer than 30 minutes \citep{2019ApJ...883...88B}.  The wide field and high sampling of the lunar telescopes will detect thousands of flares, comparable to the first results from TESS \citep{2020AJ....159...60G}, while the multi-band observations enable studies of the temperature evolution. The 10-second cadence is fast enough to detect quasi-periodic pulsations that can occur during flares, while the multi-band data will be able to characterize the formation mechanisms of these enigmatic pulsations.

\section{Hot Star Pulsations}
The field of asteroseismology studies the gravity-mode and/or pressure-mode pulsations found in certain types of stars.  These pulsations can be used to place tight constraints on fundamental stellar parameters and internal structure.  Precision asteroseismology requires a combination of high quality photometry, a fast sampling rate compared to pulsation periods, and a long baseline of observations to resolve the beat cycles between closely spaced frequencies. Hot, compact objects like subdwarf B \citep[sdB, ][]{2014MNRAS.442L..61J, 2020MNRAS.495.2844S} stars and white dwarfs \citep[WD, ][]{2017ApJS..232...23H,2019A&A...632A..42B} have been studied with optical space telescopes.  But their true interiors can be more fully explored using the UV, where the pulsation amplitudes can be an order of magnitude higher. Some work has been done to extract pulsations from archival GALEX UV data \citep{2017ApJ...845..171B,2018MNRAS.475.4768T,2019MNRAS.486.4574R}, but these efforts have been limited by the 30-minute maximum continuous baseline of the space telescope. UV observations are also crucial for tracking rapid evolution of pulsating WDs through rare dwarf novae outbursts in cataclysmic variables, where the accretion disk dominates in the optical \citep{2012ApJ...753..158S}. The uninterrupted, long baseline afforded by the lunar telescopes will result in clean power spectra to identify pulsation frequencies. The simultaneous, multi-band observations can identify the specific geometry of individual pulsation modes.

\section{Transients}
Astrophysical transients are objects experiencing drastic changes in brightness over a short timespan, including novae, supernovae (SNe), tidal disruption events (TDEs) from black holes, and kilonova counterparts to gravitational wave events. These are hot objects with blackbody temperatures peaking at $\gtrsim 10\,000~\rm{K}$, so UV observations are crucial when constraining the temperature and radius evolution. The changes in brightness within the first few minutes of these events place strong constraints on the underlying physics, but are challenging to acquire. For Type Ia SNe, the nature of the progenitor can be constrained with high cadence sampling of the light curve \citep{2015Natur.521..332O,2019arXiv190402171F}. The sample of known TDEs has increased greatly in the past decade, but many questions remain about their formation and evolution. While rare, a 24x24 deg. field results in $\sim$12 transients over a $\sim$1 week baseline, or a reasonable chance to detect one over a several hour window. This suite also serves as a test for a future facility that would cover more fields or operate for a longer period.

\end{document}